\documentclass[12pt]{iopart}
\usepackage{amssymb}
\usepackage[colorlinks=true,urlcolor=blue,linkcolor=blue]{hyperref}

\begin{document}

\title{Constructing a family of conformally flat scalar field models}
\author{Pantelis S. Apostolopoulos$^{1}$}

\begin{abstract}
Using purely geometrical methods we present a mechanism to solve the scalar
field equations of motion (non-minimally coupled with gravity) in a
spherically symmetric background. We found that the \emph{full }set of
spacetimes, which are of Petrov type O (conformally flat) and admit a \emph{%
gradient} Conformal Vector Field, can be determined completely. It is shown
that the full group of scalar field equations reduced to a \emph{single}
equation that depends only on the distance $w=r^{2}-t^{2}$ leaving the
metric function (equivalently the functional form of the scalar field or the
potential) freely chosen. Depending on the structure of the metric or the
potential $V$ (as a function of $\phi $) a solution can be found either
analytically or via numerical integration. We provide physically sound
examples and prove that (Anti)-de Sitter fits this scheme. We also
reconstruct a recently found solution \cite{Strumia:2022kez} representing an
expanding scalar bubble with metric that has a singularity and corresponds
to what is termed as Anti-de Sitter crunch.
\end{abstract}

\address{$^1$Department of Environment, Ionian University\\
Mathematical Physics and Computational Statistics Research Laboratory\\
Panagoula 29100, Island of Zakynthos, Greece}

\ead{papostol@ionio.gr}

\maketitle

\bigskip Conformal symmetries have been the subject of various studies
during the last three decades (see e.g. \cite{HallBook1}, \cite{Duggal}). In
the majority of the cases, the main reason to investigate the existence of
conformal symmetries in General Relativity was the reduction of the
complexity of the resulting system of partial differential equations (pdes)
in order to locate, more easily, an exact solution of the Einstein Field
Equations (EFEs). However, as the generality of the underlying geometry is
increased the symmetry pdes and the equation of motion become progressively
highly non-linear and often lead to models without a clear physical meaning.
On the other hand, there are sufficiently enough cases where physically
sound models admit a conformal symmetry (proper or not) that represents an
inherent constituent of their kinematical and dynamical structure. The most
well known example of this situation is the isotropic, homogeneous and
conformally flat Friedmann-Lema\^{\i}tre cosmological model which admits a
9-dimensional Lie algebra of \emph{proper} Conformal Vector Fields (CVFs) 
\cite{Maartens-Maharaj1}. In addition it has been shown that proper CVFs are
of particular interest to construct viable astrophysical models \cite%
{Herrera:1984kfa}, \cite{Maartens:1996vi}, \cite{Herrera:2022ulc} and at the
same time it has been established the significant role of self-similar
spacetimes, admitting a proper Homothetic Vector Field (HVF), since they
represent the past and future (equilibrium) states for a vast number of
evolving vacuum and $\gamma -$law perfect fluid models \cite%
{Wainwright-Ellis}, \cite{Coley-Book}.

Throughout this paper, the following conventions have been used: the
spacetime signature is assumed $\left( -,+,+,+\right) $, lower Latin letters
denote spacetime indices $a,b,...=0,1,2,3$ and we use geometrized units such
that $8\pi G=c=1$.

The existence of a CVF $\mathbf{X}$ implies that under the infinitesimal
transformation generated by $\mathbf{X}$, the spacetime metric $g_{ab}$
satisfies: 
\begin{equation}
\mathcal{L}_{\mathbf{X}}g_{ab}=2\psi g_{ab}  \label{ConformalRelation1}
\end{equation}%
where $\mathcal{L}$ is the Lie derivative along $\mathbf{X}$ and $\psi (%
\mathbf{X})$ denotes the conformal factor representing the scale deformation
of the spacetime geometry.

The above general condition (\emph{proper} CVF) specializes to a Killing
Vector Field (KVF) ($\psi (\mathbf{X})=0$), to a Homothetic Vector Field
(HVF) ($\psi (\mathbf{X})=$const.$\neq 0$) and to a Special Conformal
Killing Vector (SCKV) when $\psi _{;ab}=0$ (where "$;$" stands for the
covariant derivative w.r.t metric $g_{ab}$).

The simplest case of a spacetime geometry admitting a maximum of 15 CVFs, is
the Minkowski spacetime with metric, in Cartesian coordinates, of the form:

\begin{equation}
ds_{\mathrm{FLAT}}^{2}=-d\tau ^{2}+dx^{2}+dy^{2}+dz^{2}.  \label{Minkowski1}
\end{equation}%
The complete Lie Algebra of CVFs for the metric (\ref{Minkowski1}) has been
determined and consists of a subalgebra of 10 KVFs, 1 proper HVF and 4 SCKVs 
\cite{ChoquetBruhat}.

For the purposes of the present work, it is convenient to transform the
metric (\ref{Minkowski1}) and the CVFs in a form such that the geometry is
foliated by spherically symmetric 2d hypersurfaces i.e. with constant and
positive ($+1$) curvature. We exploit the coordinate transformation $\left(
\tau ,x,y,z\right) \hookrightarrow \left( t,r,\varphi ,\vartheta \right) $: 
\begin{equation}
\tau \left( t,r,\varphi ,\vartheta \right) =t,\quad x\left( t,r,\varphi
,\vartheta \right) =r\sin \varphi \sin \vartheta  \label{Transformation1}
\end{equation}%
\begin{equation}
y\left( t,r,\varphi ,\vartheta \right) =r\cos \varphi \sin \vartheta ,\quad
z\left( t,r,\varphi ,\vartheta \right) =r\cos \vartheta
\label{Transformation2}
\end{equation}%
and the Minkowski metric is written 
\begin{equation}
ds_{\mathrm{FLAT}}^{2}=-dt^{2}+dr^{2}+r^{2}\left( d\vartheta ^{2}+\sin
^{2}\vartheta d\varphi ^{2}\right)  \label{Minkowski2}
\end{equation}%
whereas the CVFs take the form\footnote{%
We recall that the vectors $\mathbf{X}_{1}-\mathbf{X}_{4}$ correspond to
translations, $\mathbf{X}_{5}-\mathbf{X}_{7}$ to spatial rotations, $\mathbf{%
X}_{8}-\mathbf{X}_{10}$ to spacetime rotations (boosts), $\mathbf{X}_{11}$
represents the generator of the homothety and the vectors $\mathbf{X}_{12}-%
\mathbf{X}_{15}$ are the SCKVs.}:%
\begin{eqnarray}
\mathbf{X}_{1} &=&\partial _{t},\quad \mathbf{X}_{2}=\sin \vartheta \sin
\varphi \partial _{r}+\frac{\cos \vartheta \sin \varphi }{r}\partial
_{\vartheta }+\frac{\cos \varphi }{\sin \vartheta }\partial _{\varphi }, 
\nonumber \\
&&  \nonumber \\
\mathbf{X}_{3} &=&\sin \vartheta \cos \varphi \partial _{r}+\frac{\cos
\vartheta \cos \varphi }{r}\partial _{\vartheta }-\frac{\sin \varphi }{\sin
\vartheta }\partial _{\varphi }  \nonumber \\
&&  \nonumber \\
\mathbf{X}_{4} &=&\cos \vartheta \partial _{r}-\frac{\sin \vartheta }{r}%
\partial _{\vartheta }  \label{SphericKVFs1}
\end{eqnarray}%
\begin{eqnarray}
\mathbf{X}_{5} &=&-\partial _{\varphi },\quad \mathbf{X}_{6}=-\cos \varphi
\partial _{\vartheta }+\cot \vartheta \sin \varphi \partial _{\varphi } 
\nonumber \\
&&  \nonumber \\
\mathbf{X}_{7} &=&\sin \varphi \partial _{\vartheta }+\cot \vartheta \cos
\varphi \partial _{\varphi }  \nonumber \\
&&  \nonumber \\
\mathbf{X}_{8} &=&r\sin \vartheta \sin \varphi \partial _{t}+t\sin \vartheta
\sin \varphi \partial _{r}+\frac{t\cos \vartheta \sin \varphi }{r}\partial
_{\vartheta }+\frac{t\cos \varphi }{r\sin \vartheta }\partial _{\varphi } 
\nonumber \\
&&  \nonumber \\
\mathbf{X}_{9} &=&r\sin \vartheta \cos \varphi \partial _{t}+t\sin \vartheta
\cos \varphi \partial _{r}+\frac{t\cos \vartheta \cos \varphi }{r}\partial
_{\vartheta }-\frac{t\sin \varphi }{r\sin \vartheta }\partial _{\varphi } 
\nonumber \\
&&  \nonumber \\
\mathbf{X}_{10} &=&r\cos \vartheta \partial _{t}+t\cos \vartheta \partial
_{r}-\frac{t\sin \vartheta }{r}\partial _{\vartheta }  \label{SphericKVFs2}
\end{eqnarray}%
\begin{equation}
\mathbf{X}_{11}=t\partial _{t}+r\partial _{r}  \label{HVF2}
\end{equation}%
\begin{eqnarray}
\mathbf{X}_{12} &=&\left( r^{2}+t^{2}\right) \partial _{t}+2tr\partial _{r} 
\nonumber \\
&&  \nonumber \\
\mathbf{X}_{13} &=&2rt\sin \vartheta \sin \varphi \partial _{t}+\left(
r^{2}+t^{2}\right) \sin \vartheta \sin \varphi \partial _{r}+  \nonumber \\
&&  \nonumber \\
&&+\frac{\left( t^{2}-r^{2}\right) \cos \vartheta \sin \varphi }{r}\partial
_{\vartheta }+\frac{\left( t^{2}-r^{2}\right) \cos \varphi }{r\sin \vartheta 
}\partial _{\varphi }  \nonumber \\
&&  \nonumber \\
\mathbf{X}_{14} &=&2rt\sin \vartheta \cos \varphi \partial _{t}+\left(
r^{2}+t^{2}\right) \sin \vartheta \cos \varphi \partial _{r}+  \nonumber \\
&&  \nonumber \\
&&+\frac{\left( t^{2}-r^{2}\right) \cos \vartheta \cos \varphi }{r}\partial
_{\vartheta }+\frac{\left( r^{2}-t^{2}\right) \sin \varphi }{r\sin \vartheta 
}\partial _{\varphi }  \nonumber \\
&&  \nonumber \\
\mathbf{X}_{15} &=&2rt\cos \vartheta \partial _{t}+\left( r^{2}+t^{2}\right)
\cos \vartheta \partial _{r}+\frac{\left( r^{2}-t^{2}\right) \sin \vartheta 
}{r}\partial _{\vartheta }.  \label{SCKVs2}
\end{eqnarray}%
We select the background geometry to be spherically symmetric and Petrov
type O (\emph{conformally flat}). Under these restrictions the metric of the
spacetime is: 
\begin{equation}
ds^{2}=\mathcal{C}\left( t,r\right) ^{2}\left[ -dt^{2}+dr^{2}+r^{2}\left(
d\vartheta ^{2}+\sin ^{2}\vartheta d\varphi ^{2}\right) \right]
\label{ConformallyFlatMetric1}
\end{equation}%
where $\mathcal{C}\left( t,r\right) $ is a smooth function of its arguments.

It is seen that the spacetime given in (\ref{ConformallyFlatMetric1}) shares
common features with the standard Friedmann-Lema\^{\i}tre-Robertson-Walker
(FLRW) metric like the conformal flatness and the spherically symmetric
foliation therefore can be, in principle, matched (in some appropriate
limits) with the FLRW geometry. Due to the conformal flatness, the spacetime
(\ref{ConformallyFlatMetric1}) admits a 12-dimensional Lie Algebra of \emph{%
proper} CVFs which can be used, in general, to determine the general
solution of the null geodesic equation. In fact the existence of a proper
CVF $\mathbf{X}$ implies that there is a constant of motion along null
geodesics ($n^{a}n_{a}=0$, $n_{a;b}n^{b}=0$) \cite{Maartens-Maharaj1}: 
\begin{equation}
\left( X_{a}n^{a}\right)
_{;b}n^{b}=X_{a;b}n^{a}n^{b}+X_{a}n_{;b}^{a}n^{b}=\psi g_{ab}n^{a}n^{b}=0.
\label{Constantofmotion}
\end{equation}%
The next step of the simplification setup is to demand that one of the CVFs
is a gradient vector field $X_{a}=f_{;a}$ for some smooth function $f$ which
is equivalent to demand that its bivector $F_{ab}\left( \mathbf{X}\right) =%
\frac{1}{2}\left( X_{a;b}-X_{b;a}\right) =X_{[a;b]}$ vanishes. This
heuristic assumption can be justified from the fact that FLRW spacetime also
admits the gradient CVF $\mathbf{X}_{1}$ from which the null geodesic equation is solved. From (%
\ref{HVF2}) and (\ref{ConformallyFlatMetric1}) we obtain: 
\begin{equation}
F_{ab}\left( \mathbf{X}_{11}\right) =0\Longleftrightarrow t\mathcal{C}_{,r}+r%
\mathcal{C}_{,t}=0.  \label{GradientCondition1}
\end{equation}%
The general solution of the linear pde (\ref{GradientCondition1}) implies
that the metric function depends on the quantity $r^{2}-t^{2}$ hence $%
\mathcal{C}\left( t,r\right) =\mathcal{C}\left( r^{2}-t^{2}\right) $. The
aforementioned functional structure of $\mathcal{C}\left( t,r\right) $
indicates that is also invariant under the group of spacetime rotations
(boosts) therefore we can easily deduce: 
\begin{equation}
\mathcal{L}_{\mathbf{X}_{8}}g_{ab}=\mathcal{L}_{\mathbf{X}_{9}}g_{ab}=%
\mathcal{L}_{\mathbf{X}_{10}}g_{ab}=0  \label{ExtraKVFs}
\end{equation}%
and the spacetime (\ref{ConformallyFlatMetric1}) admits (apart from the
spatial rotations that manifest the spherical symmetry of the geometry) 3
extra KVFs thus a 6-dimensional Lie Algebra of isometries.

In standard scalar field theories, non-minimally coupled with gravity, the
effective action $\mathcal{S}$ of the system without any matter
contributions has the form (without loss of generality, we have ignored
surface terms which are not alter the resulting field equations) \cite%
{Winstanley:2002jt}, \cite{Sotiriou:2008rp}: 
\begin{equation}
\mathcal{S}=\int d^{4}x\sqrt{-g}\left[ \frac{1}{2}R-\frac{1}{2}\phi
_{;a}\phi ^{;a}-\frac{1}{2}\xi R\phi ^{2}-V\left( \phi \right) \right]
\label{ScalarFieldAction1}
\end{equation}%
where $R$ is the curvature scalar, $\phi $ represents the (massless) scalar
field with potential $V\left( \phi \right) $ and $\xi $ is the coupling
constant.

Variation of (\ref{ScalarFieldAction1}) with respect to a general background
metric $g_{ab}$ gives the scalar field equations: 
\begin{eqnarray}
\left( 1-\xi \phi ^{2}\right) G_{ab} &=&T_{ab}^{\mathrm{Scalar}}\equiv
\left( 1-2\xi \right) \phi _{;a}\phi _{;b}+\left( 2\xi -\frac{1}{2}\right)
\left( \phi _{;d}\phi ^{;d}\right) g_{ab}-  \nonumber \\
&&  \nonumber \\
&&-2\xi \phi \phi _{;ab}+2\xi \phi \phi _{\,;d}^{;d}g_{ab}-Vg_{ab}
\label{EFEs1}
\end{eqnarray}%
where $G_{ab}$ is the Einstein tensor and $T_{ab}^{\mathrm{Scalar}}$ is the
effective energy-momentum tensor of the scalar field contribution.

The conservation of $T_{ab}^{\mathrm{Scalar}}$ implies the corresponding
Klein-Gordon equation: 
\begin{equation}
\phi _{\,;a}^{;a}-\xi R\phi -\frac{dV}{d\phi }=0.  \label{KleinGordon1}
\end{equation}%
Obviously, the complexity of the system of equations (\ref{EFEs1}) and (\ref%
{KleinGordon1}) does not permit us to analyze a scalar field configuration
in full generality without making some simplification assumptions.
Nevertheless we will see that the geometric setup we choose, lead to the 
\emph{complete set} of solutions with sound physical interest and,
furthemore, will verify the significance (in certain cases) of the existence
of a proper CVF.

Restricting our analysis to the spacetime (\ref{ConformallyFlatMetric1}) we
note that the isometries of the underlying geometry are inherited from the
dynamics \cite{Exact-Solutions-Book}, \cite{Eisenhart} $\mathcal{L}_{\mathbf{%
KVF}}G_{ab}=0=\mathcal{L}_{\mathbf{KVF}}T_{ab}^{\mathrm{Scalar}}$ which
implies that $\phi $ and $V\left( \phi \right) $ scale $\sim \left(
r^{2}-t^{2}\right) $. Setting $w=r^{2}-t^{2}$ the field equations (\ref%
{EFEs1}) become:%
\begin{eqnarray}
B_{\,0}^{0} &=&0=V\left( w\right) \mathcal{C}\left( w\right) ^{4}-2\mathcal{C%
}\left( w\right) ^{2}\{4r^{2}\xi \phi \left( w\right) \phi \left( w\right)
^{\prime \prime }+  \nonumber \\
&&+\phi \left( w\right) ^{\prime }\left[ \left( 4\xi
r^{2}-r^{2}-t^{2}\right) \phi \left( w\right) ^{\prime }+6\xi \phi \left(
w\right) \right] \}-  \nonumber \\
&&-4\mathcal{C}\left( w\right) \{2r^{2}\mathcal{C}\left( w\right) ^{\prime
\prime }[\xi \phi \left( w\right) ^{2}-1]+  \nonumber \\
&&+\mathcal{C}\left( w\right) ^{\prime }\left[ 2\xi \left(
r^{2}-3t^{2}\right) \phi \left( w\right) \phi \left( w\right) ^{\prime
}+3\left( \xi \phi \left( w\right) ^{2}-1\right) \right] \}+  \nonumber \\
&&+4\left( r^{2}+3t^{2}\right) \mathcal{C}\left( w\right) ^{\prime 2}\left[
\xi \phi \left( w\right) ^{2}-1\right]  \label{EFE00}
\end{eqnarray}%
\begin{eqnarray}
B_{\,1}^{0} &=&0=\mathcal{C}\left( w\right) ^{2}\left[ 2\xi \phi \left(
w\right) ^{\prime }\phi \left( w\right) ^{\prime \prime }+\left( 2\xi
-1\right) \phi \left( w\right) ^{\prime 2}\right] +  \nonumber \\
&&+2\mathcal{C}\left( w\right) \left[ \mathcal{C}\left( w\right) ^{\prime
\prime }\left( \xi \phi \left( w\right) ^{2}-1\right) -2\xi \phi \left(
w\right) \phi \left( w\right) ^{\prime }\mathcal{C}\left( w\right) ^{\prime }%
\right] +  \nonumber \\
&&+4\mathcal{C}\left( w\right) ^{\prime 2}\left[ 1-\xi \phi \left( w\right)
^{2}\right]  \label{EFE01}
\end{eqnarray}%
\begin{eqnarray}
B_{\,2}^{2} &=&0=V\left( w\right) \mathcal{C}\left( w\right) ^{4}+2\mathcal{C%
}\left( w\right) ^{2}\{4t^{2}\xi \phi \left( w\right) \phi \left( w\right)
^{\prime \prime }-  \nonumber \\
&&-\phi \left( w\right) ^{\prime }\left[ \left( t^{2}+r^{2}-4\xi
t^{2}\right) \phi \left( w\right) ^{\prime }+6\xi \phi \left( w\right) %
\right] \}+  \nonumber \\
&&+4\mathcal{C}\left( w\right) \{2t^{2}\mathcal{C}\left( w\right) ^{\prime
\prime }[\xi \phi \left( w\right) ^{2}-1]-  \nonumber \\
&&-\mathcal{C}\left( w\right) ^{\prime }\left[ 2\xi \left(
3r^{2}-t^{2}\right) \phi \left( w\right) \phi \left( w\right) ^{\prime
}+3\left( \xi \phi \left( w\right) ^{2}-1\right) \right] \}+  \nonumber \\
&&+4\left( t^{2}+3r^{2}\right) \mathcal{C}\left( w\right) ^{\prime 2}\left[
1-\xi \phi \left( w\right) ^{2}\right]  \label{EFE22}
\end{eqnarray}%
\begin{eqnarray}
B_{\,3}^{3} &=&B_{\,4}^{4}=0=V\left( w\right) \mathcal{C}\left( w\right)
^{4}-2\mathcal{C}\left( w\right) ^{2}\{4\left( r^{2}-t^{2}\right) \xi \phi
\left( w\right) \phi \left( w\right) ^{\prime \prime }+  \nonumber \\
&&+\phi \left( w\right) ^{\prime }\left[ \left( r^{2}-t^{2}\right) \left(
4\xi -1\right) \phi \left( w\right) ^{\prime }+6\xi \phi \left( w\right) %
\right] \}-  \nonumber \\
&&-4\mathcal{C}\left( w\right) \{2\left( r^{2}-t^{2}\right) \mathcal{C}%
\left( w\right) ^{\prime \prime }[\xi \phi \left( w\right) ^{2}-1]+ 
\nonumber \\
&&+\mathcal{C}\left( w\right) ^{\prime }\left[ 2\xi \left(
r^{2}-t^{2}\right) \phi \left( w\right) \phi \left( w\right) ^{\prime
}+3\left( \xi \phi \left( w\right) ^{2}-1\right) \right] \}+  \nonumber \\
&&+4\left( r^{2}-t^{2}\right) \mathcal{C}\left( w\right) ^{\prime 2}\left[
\xi \phi \left( w\right) ^{2}-1\right]  \label{EFE33}
\end{eqnarray}%
where a prime \textquotedblright $^{\prime }$\textquotedblright\ denotes
differentiation w.r.t. $w=r^{2}-t^{2}$.

From eq. (\ref{EFE01}) it follows%
\[
2\xi \phi \left( w\right) \mathcal{C}\left( w\right) ^{2}\phi \left(
w\right) ^{\prime \prime }+\left( 2\xi -1\right) \phi \left( w\right)
^{\prime 2}\mathcal{C}\left( w\right) ^{2}+ 
\]%
\begin{equation}
+2\mathcal{C}\left( w\right) \left[ \mathcal{C}\left( w\right) ^{\prime
\prime }\left( \xi \phi \left( w\right) ^{2}-1\right) -2\xi \phi \left(
w\right) \phi \left( w\right) ^{\prime }\mathcal{C}\left( w\right) ^{\prime }%
\right] +4\mathcal{C}\left( w\right) ^{\prime 2}\left( 1-\xi \phi \left(
w\right) ^{2}\right) =0.  \label{GeneratorEquation1}
\end{equation}%
Replacing $\mathcal{C}\left( w\right) ^{\prime \prime }$ to the remaining
field equations (\ref{EFE00}), (\ref{EFE22}), (\ref{EFE33}) we end up with 
\emph{only one} equation namely:%
\begin{eqnarray}
V\left( w\right) &=&2\{\mathcal{C}\left( w\right) ^{2}\phi \left( w\right)
^{\prime }\left[ w\phi \left( w\right) ^{\prime }+6\xi \phi \left( w\right) %
\right] +  \nonumber \\
&&+6\mathcal{C}\left( w\right) \mathcal{C}\left( w\right) ^{\prime }\left[
2\xi w\phi \left( w\right) ^{\prime }\phi \left( w\right) +\xi \phi \left(
w\right) ^{2}-1\right] +  \nonumber \\
&&+6w\mathcal{C}\left( w\right) ^{\prime 2}\left[ \xi \phi \left( w\right)
^{2}-1\right] \}\mathcal{C}\left( w\right) ^{-4}.  \label{Potential2}
\end{eqnarray}%
It is straightforward to prove that the pair of equations (\ref%
{GeneratorEquation1}) and (\ref{Potential2}) \emph{completely determine} the
family of solutions of the system (\ref{EFE00})-(\ref{EFE33}). We note that
(as expected), with the aid of (\ref{Potential2}), the conservation equation
(\ref{KleinGordon1}) is satisfied identically thus (\ref{Potential2})
represents a first integral of the Klein-Gordon equation.

Let us consider first, for simplicity, the case of the (Anti)-de Sitter
spacetime with metric: 
\begin{equation}
ds^{2}=\frac{c_{2}^{2}}{\left( c_{1}+w\right) ^{2}}\left[
-dt^{2}+dr^{2}+r^{2}\left( d\vartheta ^{2}+\sin ^{2}\vartheta d\varphi
^{2}\right) \right]  \label{DeSitterAntiDeSitterMetric}
\end{equation}%
where $c_{1}$ and $c_{2}>0$ are arbitrary constants and $\mathcal{C}\left(
w\right) =\frac{c_{2}}{c_{1}+w}$. We verify that 
\begin{equation}
R_{\,b}^{a}=\frac{12c_{1}}{c_{2}^{2}}\delta _{\,b}^{a}=\frac{R}{4}\delta
_{\,b}^{a}.  \label{RicciDeSitter}
\end{equation}%
Here, $R$ is the Ricci scalar and the $c_{1}$ controls the sign character of
the spacetime curvature which is directly related with the cosmological
constant of the (Anti)-de Sitter spacetime. Then, for minimally coupled
scalar field $\xi =0$, eq. (\ref{GeneratorEquation1}) implies that $\phi
\left( w\right) =$const. and the potential $V(w)=\frac{12c_{1}}{c_{2}^{2}}$.

We must note that the system of equations (\ref{GeneratorEquation1}) and (%
\ref{Potential2}), although depends (in this particular geometry) only on
the distance $w=r^{2}-t^{2}$, its solution can not be obtained always in
closed form, so that must be derived case by case (possibly numerically). In
addition the form of the potential in physically relevant models must
feature a false vacuum, as well as a true vacuum or a region in which the
potential is unbounded from below. We apply this scheme to a more
complicated setup and reconstruct a recently found solution \cite%
{Strumia:2022kez} for an expanding scalar bubble within a spherically
symmetric geometry. The metric has a singularity and corresponds to what is
termed as AdS crunch. We try solutions (up to some integration constants) of
the form: 
\begin{equation}
\mathcal{C}\left( w\right) =\left[ 1-\frac{{c_{1}^{2}}}{\left(
1+w/c_{2}^{2}\right) ^{2}}\right] ^{1/2}  \label{NikosMetricFunction1}
\end{equation}%
where $c_{1}$ and $c_{2}$ are again arbitrary constants. We observe that the
main characteristic of the spacetime (\ref{NikosMetricFunction1}) is\ the
existence of a curvature singularity that occurs when $\mathcal{C}\left(
w\right) \rightarrow 0$ as $w\rightarrow \pm c_{2}^{2}\left( c_{1}\mp
1\right) $. In addition, an apparent horizon "protects" the spacetime from
the singularity because there are two null geodesic vectors, say $n^{a}$ and 
$m^{a}$, such that $n^{a}m_{a}=-1$ (representing ingoing and outgoing null
geodesics) and $n_{;a}^{a}m_{;k}^{k}=0$ (see e.g. \cite{Hawking-EllisBook}, 
\cite{Krasinski}, \cite{Hellaby:2002nx}, \cite{Apostolopoulos:2016xnm}). We
refer the reader to \cite{Strumia:2022kez} for a discussion regarding the
structure of the spacetime (\ref{NikosMetricFunction1}).

Substituting eq. (\ref{NikosMetricFunction1}) in (\ref{GeneratorEquation1})
the general solution for the scalar field (again for $\xi =0$) reads:%
\begin{equation}
\phi \left( w\right) =c_{3}\tanh ^{-1}\frac{c_{1}}{1+w/c_{2}^{2}}
\label{ScalarFieldSolution1}
\end{equation}%
provided that $c_{3}=\sqrt{6}$. Finally, we compute the potential from eq. (%
\ref{Potential2}): 
\begin{equation}
V\left( w\right) =c_{4}\frac{c_{1}^{4}\left( 1+w/c_{2}^{2}\right) ^{-4}}{%
\left[ c_{1}^{2}\left( 1+w/c_{2}^{2}\right) ^{-2}-1\right] ^{2}}%
\mbox{ \ \ \
or \ \ \ }V\left( \phi \right) =c_{4}\sinh ^{4}\frac{\phi }{c_{3}}.
\label{Potential3}
\end{equation}%
As a final application let us consider the case of a conformally coupled ($%
\xi =1/6$) scalar field with potential proposed recently in \cite{Sotiriou:2008rp}: 
\begin{equation}
V\left( \phi \right) =-\frac{\lambda }{4}\phi ^{4}+\frac{R}{4}
\label{deSitterPotential1}
\end{equation}%
where $\lambda ,R>0$ are constants.

After standard but tedious calculations, we find that the unique exact
solution of the system of equations (\ref{GeneratorEquation1}), (\ref%
{Potential2}) and (\ref{deSitterPotential1}) is: 
\begin{equation}
\phi \left( w\right) =\frac{D_{3}\left( Rw+48\right) }{D_{3}^{2}w+288\lambda 
}.  \label{ScalarFielddeSitter4}
\end{equation}%
\begin{equation}
\mathcal{C}\left( w\right) =\frac{1}{1+\frac{R}{48}w}
\label{deSitterMetricFunction1}
\end{equation}%
where $D_{3}$ is a constant of integration. It is straightforward to show
that the spacetime (\ref{deSitterMetricFunction1}) corresponds to the de
Sitter model with constant scalar curvature $R$.

In the present article we presented a mechanism to produce compatible scalar
field spacetimes in standard gravity using geometrical methods. We observed
that, depending on the structure of a given metric function $\mathcal{C}%
\left( w\right) $ or a potential $V$ (as a function of $\phi $), a solution
can be found either analytically or via numerical integration. Although the
family of spacetimes (\ref{ConformallyFlatMetric1}), (\ref%
{GeneratorEquation1}), (\ref{Potential2}) still have some sort of simplicity
(e.g. being conformally and, in certain cases, asymptotically flat like the
spacetime (\ref{NikosMetricFunction1})) however they keep their intrinsic
generality and correspond to geometries with sound physical interest. They
also indicate an eventually \emph{close connection} between these classes of
models and the existence of a \emph{gradient} CVF, so far underestimated.
\bigskip

\noindent \textbf{\large Acknowledgments}

\noindent The author wishes to thank Nikos Tetradis for very useful
discussions.\newline

\end{document}